# Optical and mechanical design of the extreme AO coronagraphic instrument MagAO-X

Laird M. Close[1,a], Jared R. Males[a], Olivier Durney[a], Corwynn Sauve[a],
Maggie Kautz[b], Alex Hedglen[b], Lauren Schatz[b], Jennifer Lumbres[b],
Kelsey Miller[b], Kyle Van Gorkom[b], Madison Jean[b], Victor Gasho[a]

[a] Steward Observatory, University of Arizona, Tucson AZ 85721, USA;
[b] College of Optical Sciences, University of Arizona, 1630 E University Blvd, Tucson, AZ 85719.

## ABSTRACT

Here we review the current optical mechanical design of MagAO-X. The project is post-PDR and has finished the design phase. The design presented here is the baseline to which all the optics and mechanics have been fabricated. The optical/mechanical performance of this novel extreme AO design will be presented here for the first time. Some highlights of the design are: 1) a floating, but height stabilized, optical table; 2) a Woofer tweeter (2040 actuator BMC MEMS DM) design where the Woofer can be the current f/16 MagAO ASM or, more likely, fed by the facility f/11 static secondary to an ALPAO DM97 woofer; 3) 22 very compact optical mounts that have a novel locking clamp for additional thermal and vibrational stability; 4) A series of four pairs of super-polished off-axis parabolic (OAP) mirrors with a relatively wide FOV by matched OAP clocking; 5) an advanced very broadband (0.5-1.7μm) ADC design; 6) A Pyramid (PWFS), and post-coronagraphic LOWFS NCP wavefront sensor; 7) a vAPP coronagraph for starlight suppression. Currently all the OAPs have just been delivered, and all the rest of the optics are in the lab. Most of the major mechanical parts are in the lab or instrument, and alignment of the optics has occurred for some of the optics (like the PWFS) and most of the mounts. First light should be in early 2019.

## 1.0 INTRODUCTION

### 1.1 Introduction to MagAO-X

MagAO-X is a unique ExAO system in that it has been targeted for primarily doing coronagraphic science in the visible part (0.5-1.0 μm) of the spectrum. This is in contrast to many of today's ExAO systems that target the NIR (1-2.4 μm) for coronagraphy like GPI and SPHERE. MagAO leverages the excellent Las Campanas site and the slightly smaller D=6.5m size of the Magellan (Clay) telescope to allow excellent Strehls in the optical.

By use of 1700 corrected modes (at 3.7 kHz) from an advanced woofer-tweeter design we predict Strehls of ~70% at Hα (0.6563 μm) in median seeing conditions --see Males et al. 2018 (these proc.) for detailed simulations of the performance of MagAO-X.

### 1.2 Scientific Advantages to Visible AO

Despite its demanding nature, visible AO has many scientific advantages over the NIR. After all, most astronomy is done in the visible, but almost no AO science was done with λ<1μm on large 6.5-10m class telescopes until recently (Close et al. 2018; these proc.). A short list of some of the advantages of AO science in the visible compared to the NIR are:

 -- **Better science detectors** (CCDs): much lower dark current, lower readnoise (<1e- with EMCCDs), much better cosmetics (no bad pixels), ~40x more linear, and camera optics can be warm, simple, and compact.

---

[1] lclose@as.arizona.edu; phone +1 520 626 5992

-- **Much Darker skies:** the visible sky is 100-10,000x darker than the K-band sky.

-- **Strong Emission lines**: access to the primary visible recombination lines of Hydrogen (Hα 0.6563 μm) --- most of the strongest emission lines are all in the visible, and these have the best calibrated sets of astronomical diagnostics. For example, the brightest line in the NIR is Paβ some *20 times less strong* than Hα (in typical Case-B recombination conditions, T~10,000K).

-- **Off the Rayleigh-Jeans tail**: Stars have much greater range of colors in the visible (wider range color magnitudes) compared the NIR which is on the Rayleigh-Jeans tail. Moreover, visible photometry combined with the IR enables extinction and spectral types to be much better estimated.

-- **Higher spatial resolution**: The 20 mas resolution regime opens up. A visible AO system at r band (λ=0.62μm) on a 6.5m telescope has the spatial resolution of ~20 mas (with full *uv* plane coverage unlike an interferometer) that would otherwise require a 24m ELT (like the Giant Magellan Telescope) in the K-band. So visible AO can produce ELT like NIR resolutions on today's 6.5-8m class telescopes.

## 1.3 Keys to good AO Performance in the Visible: Some "point design" considerations for MagAO-X

While it is certainly clear that there are great advantages to doing AO science in the visible it is also true that there are real challenges to getting visible AO to produce even moderate Strehls on large telescopes. The biggest issue is that $r_o$ is small ~15-20 cm in the visible (since $r_o=15(\lambda/0.55)^{6/5}$ cm on 0.75" seeing site). Below we outline (in rough order of importance) the most basic requirements to have a scientifically productive visible AO system on a 6.5m sized telescope:

1. **Good 0.6" V-band Seeing Site** – Large $r_o$ (>15cm at 0.55μm) and consistency (like clear weather) is critical. In particular, low wind (long $\tau_o$ > 5ms) sites are hard to find, and so this drives loop WFS update frequencies ≥1 KHz (due in large part to the fast jet stream level winds).

→ *We have a 3.7 kHz 120x120 OCAM2 PWFS to overcome this, and our site is excellent with median 0.6" seeing.*

2. **Good DM and fast non-aliasing WFS**: need many (>500) actuators (with d<$r_{o/2}$ sampling), no illuminated "bad/stuck" actuators, need at least ≥300 well corrected modes for D>6m. Currently the pyramid WFS (PWFS) is the best for NGS science (uses the full pupil's diffraction for wavefront error measurement), hence a PWFS is preferred.

→ *We have a 2040 actuator BMC 2Kilo DM tweeter (3.5μm stroke) with only one bad "bump" over 2 actuators being controlled with up to 1700 modes by a PWFS.*

3. **Minimize all non-common path (NCP) Errors:** Stiff "piggyback" design with visible science camera well coupled to the WFS –keep complex optics (like the ADC) on the common path. Keep optical design simple and as common as possible. Limit NCP errors to less than 30 nm rms. If the NCP errors are >30 nm rms employ an extra non-common path DM to fix these errors feed by a LOWFS sensor (see Males et al. 2018; these proc).

→ *We have minimized the number of non-common optics to just 3 super-polished (λ/40 P-V; 0.1 nm roughness) flat mirrors, and one similar quality OAP. Hence, we expect less than ~30 nm of NCP, but have plans for an extra NCP DM --if needed. That extra NCP DM (baselined as another ALPAO DM-97) would be fed by a low-order WFS (LOWFS) that is directly at the coronagraphic focal plane (Miller et al. 2018 these proc).*

4. **Minimize the Low Wind Effect (LWE):** This is a wavefront error that is linked to the strong radiative cooling of the secondary support spiders in low wind. It can be mostly removed by adding a low emissivity coating so that the spiders track the temperature of the night air. Also a pyramid wavefront (PWFS) sensor seems much better at sensing the LWE errors compared to SH WFS according to Milli et al. 2018 (these proc).

→ *We never see LWE with MagAO currently, but if it becomes as issue with MagAO-X we expect the PWFS to sense it and the tweeter to eliminate it. We not expecting LWE to be a problem for MagAO-X*

5. **Minimize the Isolated Island Effect**: Unfortunately, the push towards having many corrected modes (~1700 for MagAO-X) forces visible AO systems to small subap sizes (13.5cm for MagAO-X) that approach spiders arm thicknesses (1.9 and 3.81 cm at Magellan). Hence some sub-apertures are mostly in the shadow of the spider arms and so cannot be effectively used by the WFS, allowing the DM to "run-away" in piston w.r.t. each "isolated" quadrant/section of the pupil (see Obereder et al. 2018 these proc.). This is an insidious problem which favors the use of PWFSs (which could, in theory, sense the phase difference between the quadrants in these dark zones). However, it is still unclear if PWFS can actually measure these phase differences if they are also dominated by other wavefront errors. Moreover, if the phase differences are greater than 2π the PWFS needs additional support or it will converge to the wrong (modulo 2π) solution (Esposito et al. 2017) and so an additional PWFS at another wavelength, or a real-time interferometer like a Zernike sensor (ZELDA; N'Diaye et al. 2017), or phase diversity in science focal plane; N'Diaye et al. 2018; (these proc.) must be used. This is not a solved problem on-sky.

→ *This could be a serious issue with MagAO-X if too much of our 13.5cm subaps are blocked by the shadow of the 3.8 cm spiders arms. A NIR ZELDA, or 2nd wavelength PWFS, may be needed to control the island effect.*

4. **Lab Testing:** Lots (and lots) of "end-to-end" closed loop testing with visible science camera. Alignment must be excellent and very stiff for all non-common path optics (for all observing conditions) to minimize NCP errors.

→ *In PI Jared Males' lab we have fully simulated telescope feeding MagAO-X in a clean room. Closed-loop tests will continue until into early 2019. We will ship MagAO-X back to the lab for further tests after each run in Chile.*

5. **Modeling/Design:** Well understood error budget feeding into analytical models, must at least expect ~60 nm rms WFE on-sky. Try to measure/eliminate vibrations from the telescope and environment with advanced rejection/filter techniques (eg. linear quadratic estimation (LQG) filters).

→ *PI Males has a full closed loop model of MagAO-X performance. And we have full Fresnel propagation of each optical surface to assess the coronagraphic performance of the system (see Lumbres et al. 2018 these proc).*

6 **High Quality Interaction Matrixes:** Excellent on-telescope IMATs with final/on-sky pupil. Take IMATs in partial low-order closed-loop to increase the SNR of the high order modes in the IMAT.

→ *We will be able to obtain high SNR IMATs internally with MagAO-X and test them rigorously with our turbulence simulator in the lab or even at the telescope.*

7 **IR camera simultaneous with Visible AO camera:** this is important since you achieve a 200% efficiency boost. Allows for excellent contingency in poor seeing when only NIR science is possible.

→ *the NIR is fully corrected by MagAO-X and will feed a future J (and maybe H band) science camera.*

8. **Leverage Differential Techniques for Enhanced Contrasts:** Differential techniques such as Spectra Differential Imaging (SDI) or Polarmetric Differential Imaging (PDI) are very effective in the visible, and when combined with Angular Differential Imaging (ADI) observations with Principal Component Analysis (PCA) data reduction techniques, can lead to very high contrast detections of $10^{-5}$ within 100mas (Males et al. 2016).

→ *MagAO-X has an excellent ADC, K-mirror, and pupil tracking loop which all together will enable long (~8 hour) pupil stabilized coronagraphic visits to targets. We will also increase contrast initially by carving a dark hole with a vAPP coronagraph (Otten et al. 2017). We expect $10^{-4}$ (raw 1s exposure) contrasts at 100mas at H$\alpha$ for faint R=10 mag guides stars (Males et al. 2016). Then the final boost in contrast will come from our dual camera SDI design where, say, H$\alpha$ and nearby continuum images can be imaged simultaneously. Then post-processing with KLIP/PCA can be used to calibrate and remove the PSF from each SDI science arm. The final SDI subtraction of H$\alpha$ – continuum should detect H$\alpha$ planets at 5$\sigma$ at contrasts of $10^{-5}$ - $10^{-6}$ at 100mas in ~1 hour.*

## 2.0 OPTICAL MECHANICAL DESIGN

For coronagraphy MagAO-X needs to minimize the vibrations on the optical table. We have selected a design solution where the instrument is self-contained in an enclosed floating (gravity invariant) optical table. The air dampening of the table will eliminate high frequency (>5Hz) vibrations from the telescope environment coupling into the instrument. However, a floating table might have low amplitude rocking at ~1Hz due to wind load etc. Hence, the rest of the MagAO-X's opto-mechanical systems (table, optical mounts, wheels etc.) are designed to be stiff enough to allow a <5 Hz solid body motion of the entire table without exciting jitter. This is critical for good coronagraphic performance. We have selected to have the actual height of the optical table to controlled, close-loop, by capacitive sensors mounted under the table. To our knowledge this is the first time this particular technical solution has been attempted at a telescope (although SPHERE has a somewhat similar system; Beuzit, J-L. private comm.).

Optically we have tried to minimize the size of the instrument, this lead to a 2 tiered optical layout with some folds for compactness. We needed an excellent all reflective design so that science could be done from 0.5-1.8μm. We needed 4 pairs of super-polished OAP relays to produce 4 pupils (#1 Alpao woofer pupil (13.1mm); #2 BMC DM Tweeter pupil (18.8mm); #3 PWFS modulator/coronagraphic pupil (9.0mm); and #4 the Lyot stop pupil (9.0 mm)). We also have 5 focal planes (#1 f/11 telescope FP; #2 f/16 FP; #3 f/57 FP; #4 f/69 PWFS/Coronagraphic mask FP; #5 final f/69 post-coronagraphic science FP (6x6" FOV)). The optical design is nearly perfect even over very broadband at high

airmasses thanks to our ADC design. The conceptual design of MagAO-X is shown below in a simple cartoon:

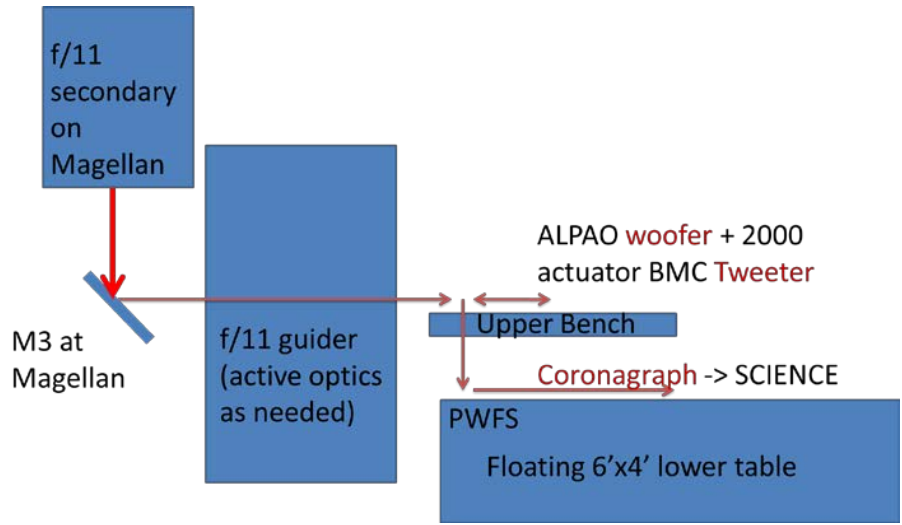

The details of the actual design can be found below in Figs. 1, 2 and 3. The optical performance is shown in Fig. 4a and 4b. Note the excellent predicted performance over 0.5-1.7μm even at high airmass.

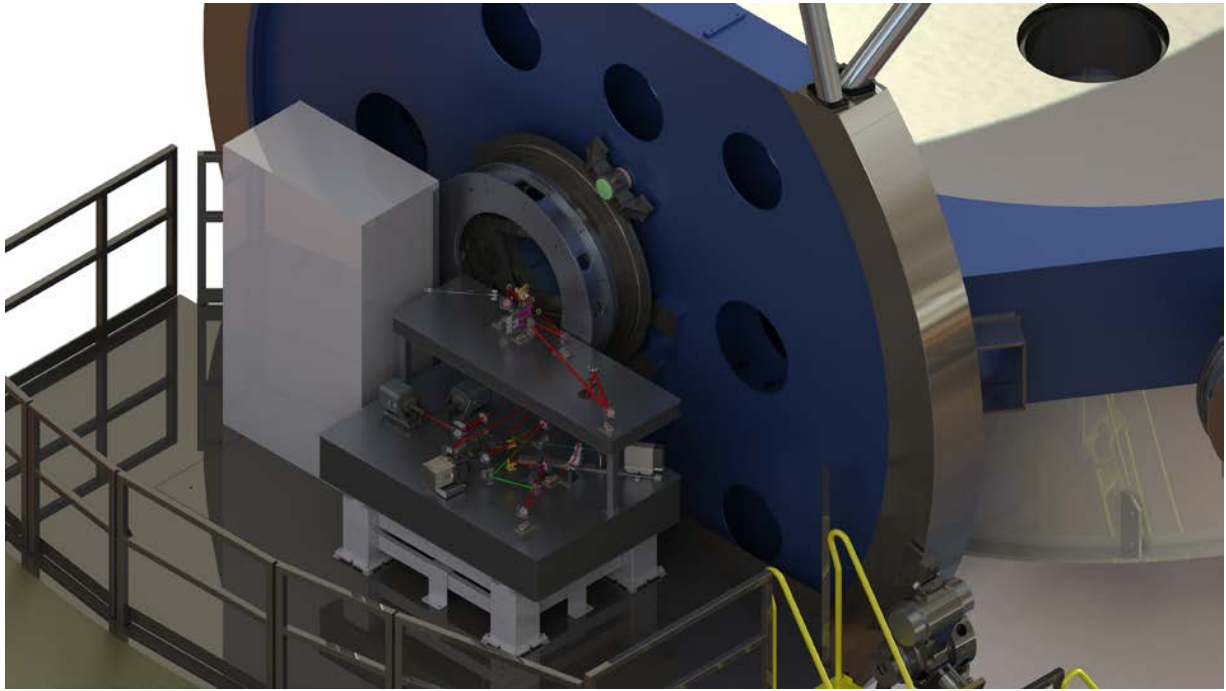

**Fig. 1:** MagAO-X as mounted at the 6.5m Magellan telescope's Nasmyth f/11 focus. The large, glycol cooled, rack to the left is for the all the MagAO-X electronics. MagAO-X is gravity invariant and mounted on a floating optical table (so neither flexure nor NCP vibrations >5 Hz are issues). Note, for clarity the dust cover is removed from the instrument. For more details see Males et al. 2018 these proc.

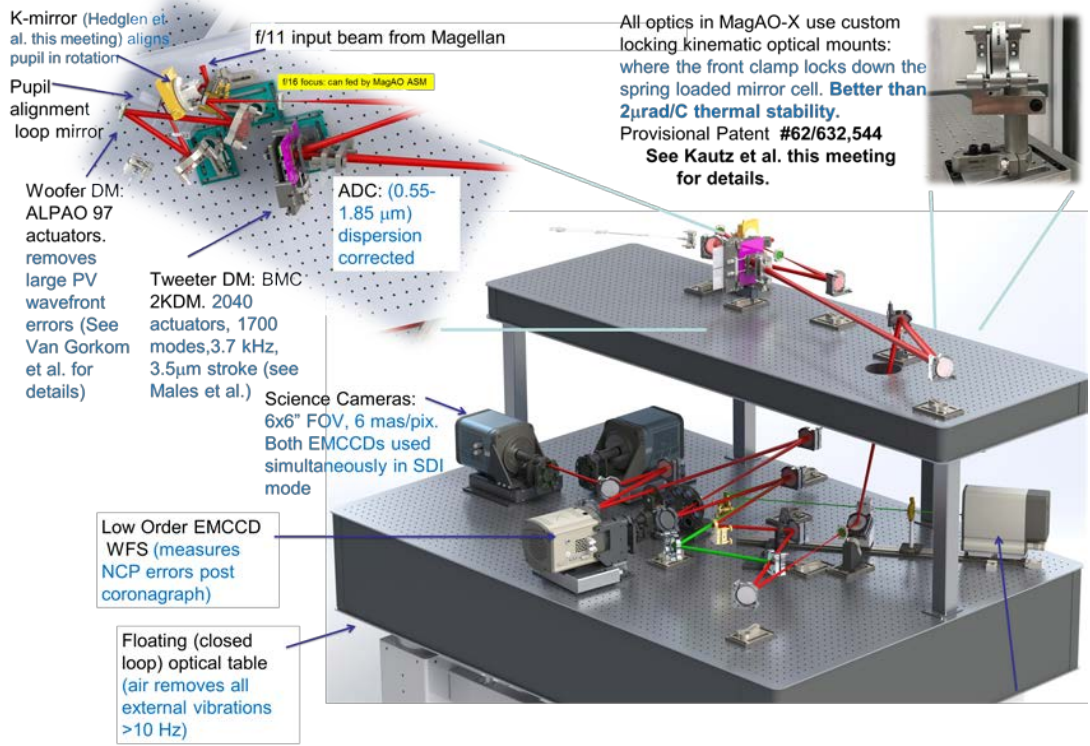

**Fig. 2:** The optical and Mechanical design for MagAO-X.

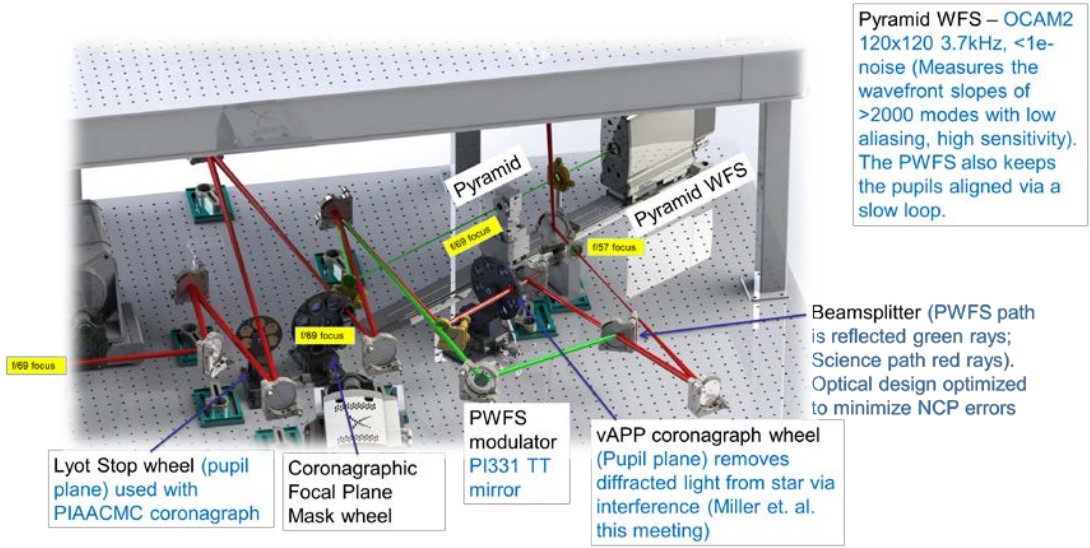

**Fig 3:** Detail of the MagAO-X vAPP Coronagraphic science (red) and PWFS (green) beam paths.

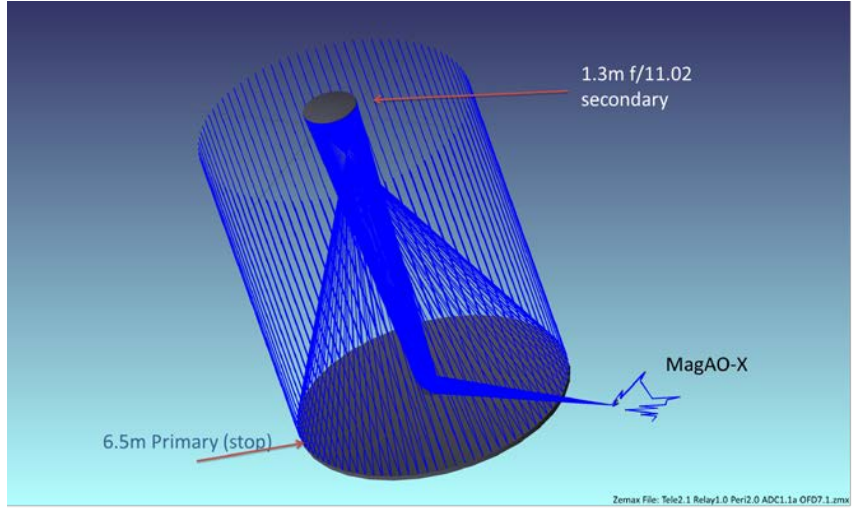

**Fig 4a:** Detail of the MagAO-X optical design.

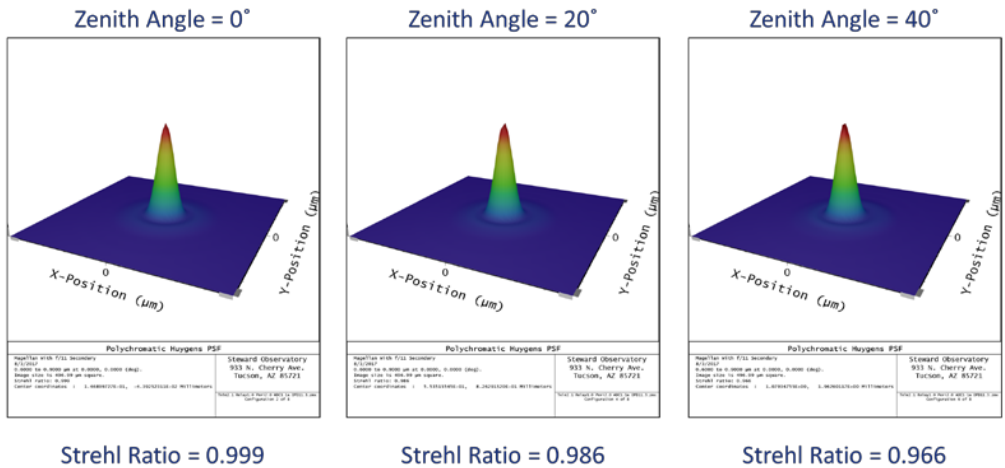

**Fig. 4b:** Full "end-to-end" optical performance of the MagAO-X optical design. In particular this design corrects atmospheric dispersion from 0.5-1.7μm with an excellent ADC. This design allows a PWFS to work broadband over the entire visible and NIR range (0.5-1.7μm) for a large range of zenith angles. Initially the MagAO-X PWFS with its OCAM2 will just work broadband over 0.55-0.95μm (shown in the polychromatic PSFs above). Of course, over any one single science filter (like r') the Strehl is >99.5%.

# 3.0 STATUS

Currently at the Extreme Wavefront Control Lab (PI Jared Males) at Steward Observatory we have fully assembled the custom floating optical table and have integrated nearly all the mounts for the optics (see Fig. 5). All the optics have been delivered in spec. We are currently in the integration, alignment and testing phase.

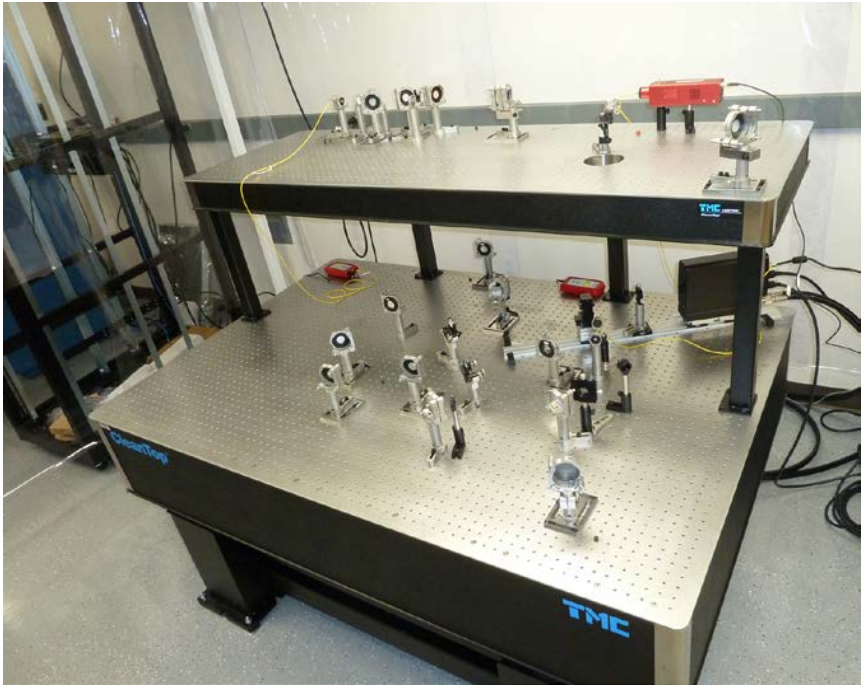

**Fig 5:** The Current view of the MagAO-X optical table in the Clean Room Steward Observatory's High-Contrast Lab

Currently we have aligned the OAPs to produce the f/69 focal plane on the tip of the pyramid (also this is the same PSF at the coronagraphic mask focal plane). See Figure 6 for an image of this PSF on our lab camera and PWFS image on the OCAM2.

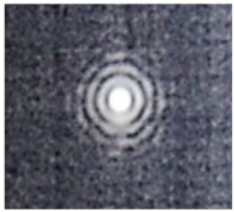
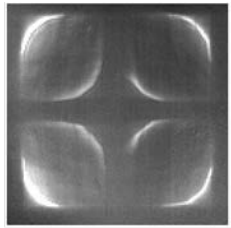

**Fig 6:** *(left)* The PSF of MagAO-X. *(Right)* the unmodulated PWFS signal on the OCAM2.

The f/69 focal plane **PSF**. This is the same PSF that exactly hits the tip the Pyramid WFS. This image suggests the initial alignment is reasonable.

The 4 Pupil Image formed by the pyramid and camera lens on the **OCAM PWFS**. This is the expected shape from an unmodulated PWFS with no wavefront error. See Schatz *et al.* for more detail.

# 4.0 NEAR FUTURE

Over summer 2018 we will finish the final alignment of our PWFS (see Schatz et al. 2018; these proceedings for more details). We will align our fabricated custom compact K-mirror to the optical table (see Hedglen et al. 2018; these proceedings for more details) and to the telescope simulator.

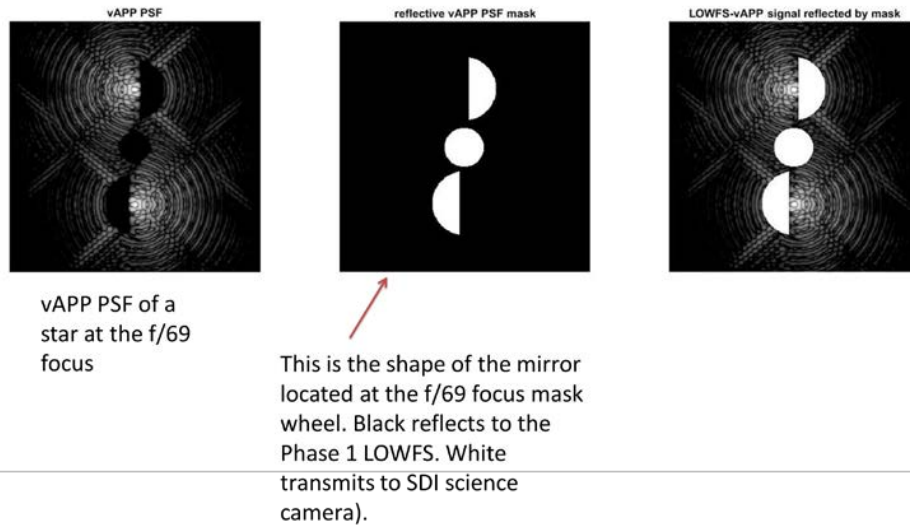

**Fig 6:** *(left)* Simulations of the vAPP coronagraphic f/69 PSF (Otten et al. 2017). Note the upper and lower "dark holes". At the coronagraphic focal plane mask wheel (see Fig. 3) we will have a mirror in the shape of the middle panel where black is the reflective surface and white is the transmissive holes (where the high contrast "planet" light passes to the SDI cameras). The right hand image is the input light to the LOWFS for post-coronagraph WFS and NCP sensing. See Miller et al. 2018 for more details. In the dark holes raw (1second exposure) contrasts of $10^{-4}$ at 100 mas should be obtained as modeled by propagation through the 22 as-built reflective optics in MagAO-X (Lumbres et al. 2018; Males et al. 2018).

The fabricated stages and filter wheels for the 2 science cameras will be integrated on the table. The three custom filter wheels (a vAPP wheel, a focal plane mask wheel, and a Lyot stop wheel) will be assembled and added to the table.

Simultaneously we will add our Alpao DM-97 woofer to the instrument (see Van Gorkom et al. 2018; these proceedings for more details). Also our custom BMC 2Kilo DM will also be integrated (see Males et al. 2018 these proceedings for more details) to the optical table. The tweeter also has a custom designed environmental control system for DM safety. By fall 2018 MagAO-X should be completely operating closed-loop in the lab.

The vAPP coronagraphic phase plate and focal plane masks are in the very final design stages (Otten et al. 2017; Miller et al. 2018 these proc.) One of our vAPPs is shown in Fig. 6 (Miller et al. 2018). Another variation of the MagAO-X vAPP would allow Zernike modal control (ZMWFS), which can allow the LOWFS to sense low order modes in the focal plane of the VAPP (Miller et al. 2018; these proceedings). The coronagraph vAPP dark hole will be maintained and enhanced by Linear Dark Field Control (LDFC). For more on our LDFC and ZMWFS lab experiments with an vAPP see Miller et al. 2018 (these proceedings).

By early 2019 the MagAO-X system should be ready to ship to Chile for first light commissioning and science. Our packing and shipping plans are well advanced and have passed a separate external review.

After the first light run we will add more advanced coronagraphs like a custom PIAACMC coronagraph to increase or contrasts at ~1-2λ/D inner working angles. However, even at first light we fully expect MagAO-X with its vAPP to be an excellent visible wavelength AO coronagraph, and to open new avenues in exoplanets and high-contrast science.

MagAO-X could not have been possible without support from the NSF's Major Research Infrastructure (MRI) research grant # *1625441 Development of a Visible Wavelength Extreme Adaptive Optics Coronagraphic Imager for the 6.5 meter Magellan Telescope* (PI: Jared Males). L.M.C.'s research was also supported by the NSF AAG program #1615408 (PI Laird Close).